\documentclass[apjl]{emulateapj}
\lefthead{TSUJIMOTO \& BEKKI}
\righthead{A Major Merger in the SMC}

\def\ltsima{$\; \buildrel < \over \sim\;$}
\def\ltsim{\lower.5ex\hbox{\ltsima}}
\def\gtsima{$\; \buildrel > \over\sim \;$}
\def\gtsim{\lower.5ex\hbox{\gtsima}}
\def\ms{$M_{\odot}$ }
\def\msp{$M_{\odot}$}

\slugcomment{Accepted for publication in ApJ Letters}

\begin{document}
\title{Chemical Signature of a Major Merger in the Early Formation of \\ Small Magellanic Cloud}

\author{Takuji Tsujimoto$^1$ and Kenji Bekki$^2$}

\affil{$^1$National Astronomical Observatory, Mitaka-shi,
Tokyo 181-8588, Japan; taku.tsujimoto@nao.ac.jp}
\affil{$^2$School of Physics, University of New South Wales, Sydney 2052, NSW, Australia}

\begin{abstract}
The formation history of the Small Magellanic cloud (SMC) is unraveled based on the results of our new chemical evolution models constructed for the SMC, highlighting the observed anomaly in the  age-metallicity relation for star clusters in the SMC. We first propose that evidence of a major merger is imprinted in the age-metallicity  relation as a dip in [Fe/H].  Our models predict that the major merger with a mass ratio of 1:1 to 1:4 occurred at $\sim$7.5 Gyr ago, with a good reproduction of the abundance distribution function of field stars in the SMC. Furthermore, our models predict a relatively large scatter in [Mg/Fe] for $-1.4 \le {\rm [Fe/H]} -1.1$ as a reflection of a looping feature resulting from the temporally inverse progress of chemical enrichment, which can be tested against future observational results. Given that the observed velocity dispersion ($\sim 30$ km s$^{-1}$) of the SMC is much smaller than that ($\sim 160$ km s$^{-1}$) of the Galactic halo, our finding strongly implies that the predicted merger event happened in a small group environment that was far from the Galaxy and contained a number of small gas-rich dwarfs comparable to the SMC. This theoretical view is extensively discussed in the framework that considers a connection with the formation history of the Large Magellanic cloud.  
\end{abstract}

\keywords{Magellanic Clouds --- galaxies: formation --- galaxies: star clusters --- stars: abundances}

\section{Introduction}

Merging plays a key role in the hierarchical galaxy formation in the cold dark matter (CDM) Universe \citep{White_78, White_91}. Numerical simulations based on the CDM scenario demonstrate how the galaxies build up hierarchically \citep[e.g.,][]{Abadi_03, Springel_05}. In addition to the theoretical ground, the observed mass function of galaxies as well as that for cluster of galaxies have evolved with redshifts, suggestive of the merging process over the cosmic time, though some discrepancy between the prediction and the observation exists \citep[e.g.,][]{Pozzetti_07, Marchesini_08, Bahcall_03}. Furthermore, some fraction of galaxies at all redshifts show an ongoing merger. For high-z galaxies, we witness an asymmetric structure composed of many blobs, indicative of a major merger \citep{Conselice_03}. In the near field, recent minor merger can be seen as a tidal stream in the halo such as the Sagittarius stream in the Galaxy \citep{Ibata_94} or the Giant Southern Stream in M31 \citep{Ibata_01}.

Recent studies have suggested that Magellanic clouds (MCs) are the promising candidates to exhibit a sign of past merger. First, \citet{Bekki_08} have claimed that kinematical difference between H I gas and older stars in the Small MC (SMC) \citep{Stanimirovic_04, Harris_06} is an end result of a major merger occurred in the early stage of the SMC formation. Secondly, the Large MC (LMC) is found to possess an extended stellar halo exhibiting a broad metallicity spread, which can be linked with the nature of hierarchical formation through mergers \citep{Majewski_09}. Finally, the LMC and SMC might have belonged to a larger Magellanic group \citep{D'Onghia_08}, which ensures a gas infall onto the MCs owing to the smaller velocity dispersion inside the MC group.  Note that such an infall is hard to realize once the MCs become the Galaxy satellites like the present time.

Chemical features of MCs have been extensively investigated through the obserations for young and old stars belonging to the field as well as star clusters \citep[see the review by][]{Hill_04}, though there are less studies for the SMC where its past information can be obtained only through the star clusters.  The merit of clusters as a tool to trace the chemical evolution is that their ages estimated from the color-magnitude diagram are far more precise than those of field stars. Accordingly, how the metallicity evolved with time, i.e., the age-metallicity relation of star clusters, deserves the first-ranked information for the study on the history of MCs. The past ravaging event such as a major merger, if occurred in galaxies, will imprint some signature in their age-metallicity relation.

In the general scheme on chemical evolution of galaxies, the metallicity gradually increases with time. If we spot a break in this relation for a galaxy, what does this feature tell us about the formation history of the galaxy?  In this letter, we report that there exists a clear dip in [Fe/H] in the age-metallicity relation among the star clusters in the SMC, and conclude that it is a relic of a major merger occurred at the early stage of SMC formation. Its implication for the history of SMC in connection with the LMC and the Galaxy is also discussed.

\section{Evidence and model}

\subsection{A dip in [Fe/H]}

\begin{figure}[t]
\vspace{0.2cm}
\begin{center}
\includegraphics[width=8cm,clip=true]{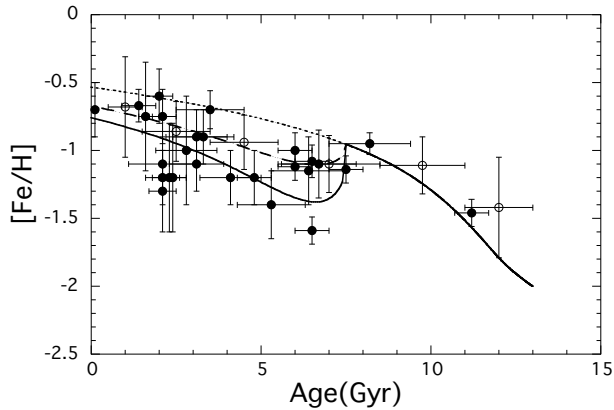}
\end{center}
\vspace{0.3cm}
\caption{Comparison of age-metallicity relation between the models and the observations. Three curves represent the results of the model in which the merger with the mass ratio of one to four is included (solid curve), the model with an equal mass merger (dashed curve), and the model with no merger event (dotted curve), respectively. Filled circles denote the observed data of star clusters, compiled from various papers (Da Costa \& Hatzidimitriou 1998; Piatti et al. 2001, 2005, 2007a,b; Kayser et al. 2007; Glatt et al. 2008a,b), whereas open circles show the mean metallicity in six age bins for 386 field SMC stars \citep{Carrera_08a}.}
\end{figure}

Filled circles in Figure 1 denote the correlation of [Fe/H] with age for individual star clusters. Here we have compiled the latest data from various references, and the result for 386 field stars that are divided into six age bins is also shown (open circles). A fairly good agreement between clusters and field stars is seen. As an overall feature, we see a gradually increasing trend of [Fe/H] with time. At the same time, a gap in level between the evolutionary path for $t >$8 Gyr and that for $t<$7 Gyr is conspicuous. In other words, the Fe abundance once decreases $\sim$8 Gyr ago and then restarts to increase around $t$=7 Gyr till the present. Here we want to stress that this dip is not an artifact resulting from the systematic difference between the multiple studies. Except for one cluster \citep[at $t$=3.5 Gyr,][]{Costa_98}, ages and metallicities of other clusters are the results of two groups by Glatt's et al. and Piatti et al., each of which occupies the separate age-range in the plot, i.e., for old ($t>$ 6 Gyr) clusters, and young and intermediate-age clusters, respectively. Thus, the observed dip is established by one study. Moreover, the previous results on old clusters are also suggestive of the presence of dip. Figure 2 shows the data of Glatt's group extracted from Fig.~1 for the clusters with their ages older than 5 Gyr (black circles),  together with the results by \citet{Costa_98} (red circles) and by \citet{Mighell_98} (blue circles). We see the dip convincingly for one while marginally for the other. Meanwhile, this figure makes the inhomogeneity of metallicity values around $t\sim 6$-6.5 Gyr by Glatt's group stand out. The lower value of [Fe/H] for one cluster seems compatible with other two studies. Finally, it should be of note that the dip feature is supported by the data of field stars as well as by an increasing rate of [Fe/H] for young and intermediate-age clusters as shown in \S 3.

Only a mechanism to make such a dip in [Fe/H] is considered to be an external dilution of the interstellar matter (ISM) by an accretion of very low-metallicity gas. The degree of dilution is not small: the decrease in [Fe/H] is estimated to be 0.3 dex at least. That is equivalent to an accretion of pristine gas with an equal mass to the existing ISM at that time, likely an order of $10^8$\msp, as the least mass scale. Therefore, the observed decrease in [Fe/H] in the SMC has a strong implication for the past occurrence of a major (not minor) merger.

In the following, we construct the model for chemical evolution of the SMC, based on the major merger scenario.

\begin{figure}[t]
\vspace{0.2cm}
\begin{center}
\includegraphics[width=6cm,clip=true]{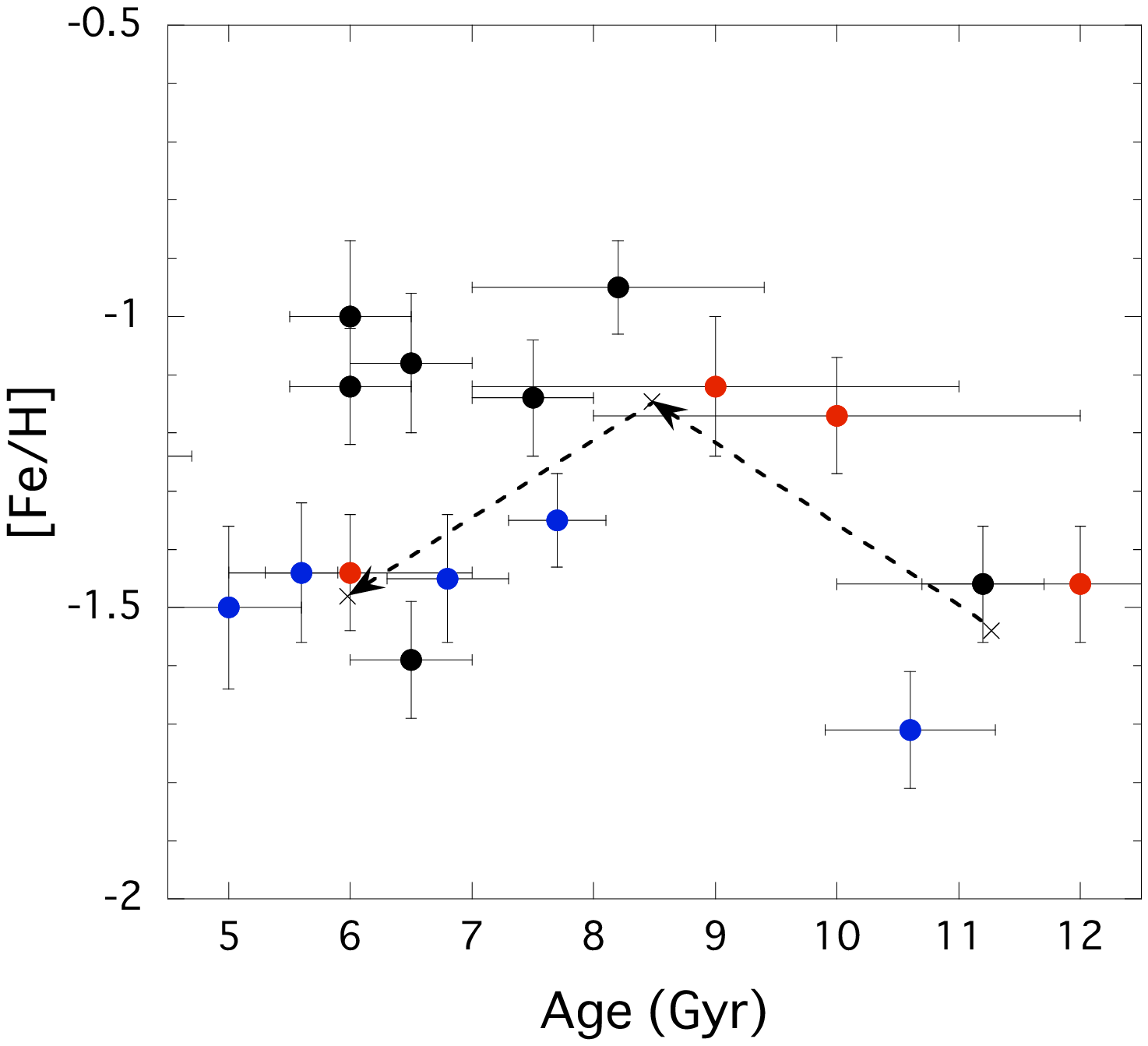}
\end{center}
\vspace{0.3cm}
\caption{Age-metallicity relation for old ($ t >5$ Gyr) clusters by three different groups. Black circles represent the data of Glatt's group extracted from Fig.~1 for the clusters with their ages older than 5 Gyr. Red and blue circles are taken from \citet{Costa_98} and \citet{Mighell_98}, respectively. The overall trend implied by the age-binned averages (crosses; $t \leq$ 7 Gyr, 7 $< t \leq$ 10 Gyr, $t>$ 10 Gyr) is shown by the arrows. Here we neglect three data situated around ($t$, [Fe/H])=(6, -1.1) (see the text).}
\end{figure}

\subsection{Model}

Our prime aim is to reproduce the age-metallicity relation of the SMC, by incorporating the major merger phenomenon into the model. Let $\psi(t)$ be the star formation rate and $A(t)$, $A'(t)$ be the gas infall rate, the accretion rate by a major merger, respectively,  then the gas fraction $f_g(t)$ and the abundance of heavy-element $i$ $Z_i(t)$ in the gas change with time according to 
\begin{equation}
\frac{df_g}{dt}=-\alpha\psi(t)+A(t)+A'(t)
\end{equation}
\begin{eqnarray}
\frac{d(Z_if_g)}{dt}=-\alpha Z_i(t)\psi(t)+Z_{A,i}(t)A(t)+y_{{\rm II},i}\psi(t) \nonumber \\
 +y_{{\rm Ia},i}\int^t_0 \psi(t-t_{\rm Ia})g(t_{\rm Ia})dt_{\rm Ia}+Z_{A',i}(t)A'(t) \ \ ,
\end{eqnarray}
\noindent where $\alpha$ is the mass fraction locked up in dead stellar remnant and long-lived stars, $y_i$ is the heavy-element yield from an SNII or SN Ia, and $Z_{A,i}$ ($Z_{A',i}$) is the abundance of heavy element  contained in the infalling gas (the accreting gas driven by a merger). 

The star formation rate $\psi(t)$ is assumed to be proportional to the gas fraction with a constant rate coefficient of 0.03 Gyr$^{-1}$. For the infall rate, we adopt the formula that is proportional to $\exp(-t/\tau_{\rm in})$ with a timescale of infall $\tau_{\rm in}$. In this formula, $\tau_{\rm in}$= 0.3 Gyr is adopted, and the metallicity $Z_{A,i}$ of an infall is assumed to be [Fe/H]=-2 with a SN-II like enhanced [$\alpha$/Fe] ratio. For reference, the values for the corresponding rate and timescale in the solar neighborhood are predicted to be 0.4 Gyr$^{-1}$ and 5 Gyr, respectively \citep{Tsujimoto_07}.

The effect of a merger on chemical evolution is devised to be an additional term $A'(t)$ which has the same form as $A(t)$. Here, the onset of $A'(t)$ is set at $t=5.5$ Gyr with the assumption that the age of SMC is 13 Gyr old. Thus, $A'(t)=0 \ {\rm for} \ t<5.5 \ {\rm Gyr}, A'(t)\propto \exp(-(t-5.5)/\tau_{\rm in}) \ {\rm for}\  t\geq5.5 \ {\rm Gyr}$. The timescale of accretion is assumed to be $\tau_{\rm in}$= 1 Gyr. This somewhat long timescale is adopted to avoid the sudden decrease in [Fe/H] by a merger. The total gas accretion is normalized so that $\int^\infty_0 (A(t)+A'(t)) dt = 1$. The mass ratio between two functions is the crucial factor to control the impact of the merging. We calculate three cases of $A: A' =1:0, 1:1, 1:4$. In all cases, the metallicity $Z_{A'}$ is set to be [Fe/H]=-2 so as to produce the dip in [Fe/H] by a merger event.

For the initial mass function, we assume a power-law mass spectrum with a slope of -1.35, which is combined with the nucleosynthesis yields of SNe Ia and II taken from \citet{Tsujimoto_95} to deduce $y_{{\rm II},i}$ and $y_{{\rm Ia},i}$. We apply an instantaneous recycling approximation to SN II progenitors, leading to the ejection rate of heavy elements, which is proportional to $\psi(t)$. In contrast, the effect of time delay yields the ejection rate from SN Ia's, which is proportional to $\psi(t-t_{\rm Ia})$. It is assumed that $t_{\rm Ia}$ spans over some range according to a distribution function $g(t_{\rm Ia})$.  Here we adopt that the fraction of the stars that eventually produce SNe Ia for $3-8$\ms is 0.05 with a box-shaped $g(t_{\rm Ia})$ for $0.5\leq t_{ Ia}\leq3$ Gyr \citep{Yoshii_96}. 

\section{Results}

\begin{figure}[t]
\vspace{0.2cm}
\begin{center}
\includegraphics[width=7.2cm,clip=true]{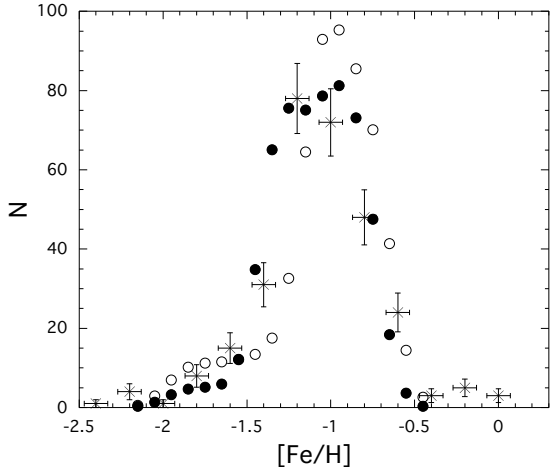}
\end{center}
\vspace{0.3cm}
\caption{Observed and predicted abundance distribution functions of SMC stars against the Fe abundance. The observed distribution represented by crosses with error bars is obtained from the selection of data taken from \citet{Carrera_08a}. See the text for detail. The model distribution denoted by filled circles is the one calculated under the assumption of the merging with the mass ratio of one to four. The other one (open circles) is with the one-to-one merger model. These distributions are convolved using the Gaussian with a dispersion of 0.1 dex in [Fe/H], and are normalized to coincide with the total number of the sample stars of observation.
}
\end{figure}

First, we show the calculated age-metallicity relations for three cases (dotted curve: no merger, dashed curve: one-to-one merger, solid curve: one-to-four merger), together with the observations. Here ages represent the lookback time. The merger models can reproduce not only the observed dip around 7 Gyr ago but also the increasing rate of [Fe/H] before and after the dip. There are several clusters deviating from the evolutionary paths predicted by the merger models. These are classified into two groups; one is well fitted by the model with no merger (two clusters) and the other is composed of ones dropping below the curves due to their lower metallicities than the predictions at each age (the assemblage of four clusters at $t \sim 2$ Gyr and one at $t \sim 6.5$ Gyr). The presence of such clusters might imply that there existed regions where a gas accretion driven by a major merger did not take place, and that additional accretion events, possibly by minor mergers, locally diluted the ISM in the SMC, along with another possibility that low-metallicity clusters came into the SMC from other galaxies through major/minor mergers.

The successful merger models should be verified by checking whether they are compatible with other chemical features of SMC. In general, stellar abundance distribution function (ADF) gives a stringent constraint on the models. For the SMC, from over 350 field stars \citep{Carrera_08a}, the observed ADF can be constructed. The distribution denoted by crosses in Figure 2 shows the derived ADF form their data, in which we exclude stars in three outer fields where the metallicity on average is distinctly lower  than other fields. In the end, 293 stars are entered. The ADF thus obtained is compared with the results by the one-to-four merging model (filled circles) and the one-to-one model (open circles). A fairly good agreement with the observation, especially for the one-to-four model, can be seen. 

\begin{figure}[t]
\vspace{0.2cm}
\begin{center}
\includegraphics[width=7cm,clip=true]{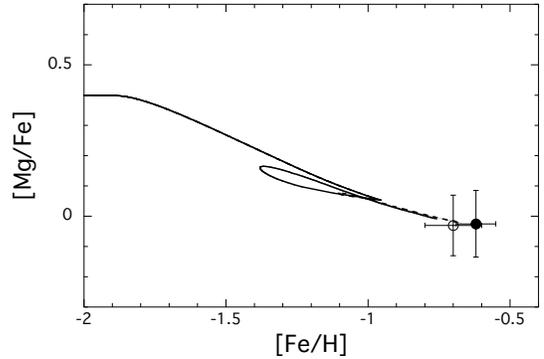}
\end{center}
\vspace{0.3cm}
\caption{Predicted correlations of [Mg/Fe] with [Fe/H] for SMC stars, together with the observed data at present (filled circle: Venn 1999, open circle: Rolleston et al. 2003). Solid and dashed curves are the same as for Fig.~1.}
\end{figure}

Finally, we show the resultant correlation of [Mg/Fe] with [Fe/H] for two merging models. After drawing a  loop which is a result of the increase-decrease-increase chain of [Fe/H] induced by a merger, both curves reach the present-day observed value. For the present value, chemical compositions for A-type supergiant \citep{Venn_99} and B-type dwarf \citep{Rolleston_03} are shown. Our merging models predict a relatively large scatter in [Mg/Fe] for the range of -1.4\ltsim [Fe/H] \ltsim -1.1 as a reflection of a looping feature.

\section{Discussion and Conclusions}

Motivated by the observed dip in [Fe/H] seen in the age-metallicity relation for the star clusters in the SMC, we construct the chemical evolution model in which the gas infall driven by a major merger is 
incorporated. Accordingly, we claim that the SMC experienced a major merger in the manner that two progenitor  galaxies of the SMC merged with the mass ratio of 1:1 to 1:4 around 7.5 Gyr ago. Different speed of chemical enrichment between two galaxies results in creating the gap/discontinuity of age-[Fe/H] relation as observed. \citet{Harris_04} found that the SMC formed relatively few stars between 8.4 and 3 Gyr ago and thus did not claim the presence of a starburst population around 7.5 Gyr ago. This together with our finding implies that a major merger occurred in the SMC did not trigger a major starburst for some physical reasons but proceed with a moderate star formation. We do not doubt that 
the detailed processes involved in SMC evolution, including the connection between the potential merger events and the star formation history, will be unraveled  by the future databases of detailed elemental abundances of old stellar population in the SMC, which will be brought by the new-generation of surveys such as HERMES.

Here we have proposed a galaxy-galaxy merger in the framework that the SMC merged with a gas-rich dwarf galaxy including a stellar component, since such process would change the dynamical properties of the SMC, the signature of which is suggested to be imprinted in the kinematical difference between H I gas and old stars \citep{Bekki_08}. If we put aside the kinematical discussion, an accretion event of a massive HI gas cloud with a mass of $10^8$ \ms order, which is not associated with a galaxy embedded in a dark matter halo, also explains the observed [Fe/H] dip in the SMC. However, such a massive gas accretion event is almost outside the range of possibility. That's firstly because such massive isolated gas clouds are extremely rare and usually associated with tidal debris as a result of interacting galaxies \citep[e.g.,][]{Koribalski_04}. Nowhere in the Local Group has possessed a potentiality to provide such a huge amount of gas with a short timescale of $\sim 1$Gyr. On the other hand, if the SMC was already within the Galactic halo about 7.5 Gyr ago, a gas accretion onto the SMC rather than onto the Galaxy would be severely suppressed owing to the large relative-velocity between the SMC and the cloud in the Galactic halo. In the end, it is concluded that the observed dip is attributable to the gas infall associated with a galaxy-galaxy merger occurred to the SMC.
 
The difficulty of a gas accretion onto the SMC inside the Galaxy discussed above also holds for a merger case. Given that the observed velocity dispersion \citep[$\sim 30$ km s$^{-1}$,][]{Harris_06} of the SMC is much smaller than that \citep[$\sim 160$ km s$^{-1}$ estimated from $\sqrt{2}V_c$,][]{Binney_87} of the Galactic halo, the major  merger is highly unlikely to happen within the Galactic halo. Accordingly, we suggest the following two possible schemes on the SMC's merger event. One is that the merger event could have happened in a small group where  (i) the velocity dispersion was comparable to that of the SMC and (ii) there was a number of gas-rich dwarfs. Recent cosmological simulations have shown that the Galaxy grows through accretion of small groups \citep[e.g.,][]{Li_08}, which is compatible with the view that the small group including the SMC might have existed before and accreted into the Galaxy potential to lose other member galaxies in the Galactic halo. The other is that the SMC had been  a binary system of two gas-rich dwarfs until $\sim$7.5 Gyr ago and then merged with each other to form a single dwarf with an extended HI gas disk.

By incorporating the hypothesis that the MCs have a common diffuse dark halo with the mass larger than $\sim 2 \times 10^{10} M_{\odot}$ into the orbital models for the MCs, \citet{Bekki_08a} succeeded in reproducing the present 3D velocities of MCs, which are consistent with the latest proper motion
measurements \citep{Kallivayalil_06}. With the same model, he found that the strong tidal interactions among the LMC, the SMC, and the Galaxy for the last 2 Gyr led to the formation of the Magellanic Stream. It is, however, unclear how the possible common halo was formed in the histories of the LMC and the SMC. As the possible mechanism to form the common halo, it can be regarded as the remnant structure evolved from a group where both the LMC and the SMC had belonged until recently. This scenario leads to a plausible picture that such a group provided the SMC with the major merger event as proposed in the present study. We note that the recent dynamical coupling of the LMC and the SMC around 3-4 Gyr ago is another promising mechanism for the formation of the common halo in terms of the origin  of ``the age gap'' in the globular cluster system  of the LMC \citep{Bekki_04}.

Interestingly, we see a similar dip in [Fe/H] at $\sim$8-9 Gyr ago for field stars in the LMC \citep{Carrera_08b}, though in contrast to the continuous age-distribution of field stars, the LMC star clusters show the lack of the intermediate-age (4 $< t <$ 9 Gyr) clusters \citep[e.g.,][]{Olszewski_91, Dirsch_00}. Whether this observed dip in [Fe/H] for the LMC has some connection with that for the SMC at $\sim$7.5  Gyr ago is the issue to be solved with a high priority. If the LMC's dip is due to a  merger event occurred at the slightly older age than the SMC, the event will imprint some observable records in stellar structure and kinematics of the present-day LMC. The potential candidates are considered to be the observed extended stellar halo \citep[e.g.,][]{Majewski_09} and the presence of the thick disk \citep[e.g.,][]{vandermarel_02}. In the end, we can bear the following picture in mind. A  small group, so-called the Magellanic group, contained both the LMC and the SMC at least until the completion of a merging process about 6.5 Gyr ago, and then it had been partly or completely disrupted by the Galactic tidal field during its accretion onto the Galaxy. It results in the present situation that the other group member galaxies exist as satellite galaxies with orbits similar to the MCs in the Local Group \citep{D'Onghia_08}. Validation of this scheme together with the issue of which satellite galaxies originally belonged to the Magellanic group will be discussed in our future papers. 

\acknowledgements
The authors  thank the anonymous referee for useful comments that helped improve this paper. This work is assisted in part by Grant-in-Aid for Scientific Research (21540246) of the Japanese Ministry of Education, Culture, Sports, Science and Technology.


\begin{thebibliography}{}
\bibitem[Abadi et al.(2003)]{Abadi_03}
Abadi, M. G., Navarro, J. F., Steinmetz, M., \& Eke, V. R. 2003, ApJ, 591, 499
\bibitem[Bahcall \& Bode(2003)]{Bahcall_03}
Bahcall, N. A., \& Bode, P. 2003, ApJ, 588, L1
\bibitem[Bekki (2008)]{Bekki_08a}
Bekki, K. 2008, ApJ,  684, L87
\bibitem[Bekki \& Chiba(2008)]{Bekki_08}
Bekki, K., \& Chiba, M. 2008, ApJ, 679, L89
\bibitem[Bekki et al. (2004)]{Bekki_04}
Bekki, K., Couch, W. J., Beasley, M. A., Forbes, D. A., Chiba,
M., Da Costa, G. S. 2004, ApJ,  610, 93
\bibitem[Binney \& Tremaine(1987)]{Binney_87}
Binney, J., \& Tremaine, S. 1987, Galactic dynamics (Princeton: Princeton Univ. Press)
\bibitem[Carrera et al.(2008a)]{Carrera_08a}
Carrera, R., Gallart, C., Aparicio, A., Costa, E., M\'{e}ndez, R. A., \& No\"{e}l, N. E. D. 2008a, AJ, 136, 1039
\bibitem[Carrera et al.(2008b)]{Carrera_08b}
Carrera, R., Gallart, C., Hardy, E, Aparicio, A., \& Zinn, R. 2008b, AJ, 135, 836
\bibitem[Conselice et al.(2003)]{Conselice_03}
Conselice, C. J., Bershady, M. A., Dickinson, M., \& Papovich, C. 2003, AJ, 126, 1183
\bibitem[Da Costa \& Hatzidimitriou(1998)]{Costa_98}
Da Costa, G. S., \& Hatzidimitriou, D. 1998, AJ, 115, 1934
\bibitem[Dirsch et al.(2000)]{Dirsch_00}
Dirsch, B., Richtler, T., Gieren, W. P., \& Hilker, M. 2000, A\&A, 360, 133
\bibitem[D'Onghia \& Lake(2008)]{D'Onghia_08}
D'Onghia, E., \& Lake, G. 2008, ApJ, 686, L61
\bibitem[Glatt et al.(2008a)]{Glatt_08a}
Glatt, K. et al. 2008a, AJ, 135, 1106
\bibitem[Glatt et al.(2008b)]{Glatt_08b}
Glatt, K. et al. 2008b, AJ, 136, 1703
\bibitem[Harris \& Zaritsky(2004)]{Harris_04}
Harris, J., \& Zaritsky, D. 2004, AJ, 127, 1531
\bibitem[Harris \& Zaritsky(2006)]{Harris_06}
Harris, J., \& Zaritsky, D. 2006, AJ, 131, 2514
\bibitem[Hill (2004)]{Hill_04}
Hill, V. 2004, in Carnegie Observatories Astrophysics Series, Vol.~4: Origin and Evolution of the Elements, ed. A. McWilliam \& M. Rauch (Cambridge: Cambridge Univ. Press), p. 203
\bibitem[Ibata et al.(1994)]{Ibata_94}
Ibata, R. A., Gilmore, G., \& Irwin, M. J. 1994, Nature, 370, 194
\bibitem[Ibata et al.(2001)]{Ibata_01}
Ibata, R. A. et al. 2001, Nature, 412, 49
\bibitem[Kallivayalil et al. (2006)]{Kallivayalil_06}
Kallivayalil, N., van der Marel, R. P., \& Alcock, C. 2006, ApJ, 652, 1213
\bibitem[Kayser et al.(2007)]{Kayser_07}
Kayser, A. et al. 2007, in Stellar Populations as Building Blocks of Galaxies, IAU Symp.~241, ed. A. Vazdekis \& R. F. Peletier (Cambridge: Cambridge Univ. Press), p. 351
\bibitem[Koribalski  et al.(2004)]{Koribalski_04}
Koribalski, B. S. et al. 2004, AJ, 128, 16
\bibitem[Li \& Helmi(2008)]{Li_08}
Li, Y.-S., \& Helmi, A. 2008, MNRAS, 385, 1365
\bibitem[Majewski et al.(2009)]{Majewski_09}
Majewski, S. R., Nidever, D. L., Mu\~{n}oz, R. R., Patterson, R. J., Kunkel, W. E., \& Carlin, J. L. 2009, in The Magellanic System: Stars, Gas, and Galaxies, IAU Symp.~256, ed. J. Th. van Loon \& J. M. Oliveira  (Cambridge: Cambridge Univ. Press), p. 51
\bibitem[Marchesini et al.(2008)]{Marchesini_08}
Marchesini, D. et al. 2008, submitted to ApJ
\bibitem[Mighell et al.(1998)]{Mighell_98}
Mighell, K., Sarajedini, A., \& French, R. S. 1998, AJ, 116, 2395
\bibitem[Olszewski et al.(1991)]{Olszewski_91}
Olszewski, E. W., Schommer, R. A., Suntzeff, N. B., \& Harris, H. C. 1991, AJ, 101, 515
\bibitem[Piatti et al.(2001)]{Piatti_01}
Piatti, A. E., Santos, J. F. C. Jr, Clari\'{a}, J. J., Bica, E., Sarajedini, A., \& Geisler, D. 2001, MNRAS, 325, 792
\bibitem[Piatti et al.(2005)]{Piatti_05}
Piatti, A. E., Sarajedini, A., Geisler, D., Seguel, J., \& Clark, D. 2005, MNRAS, 358, 1215
\bibitem[Piatti et al.(2007a)]{Piatti_07a}
Piatti, A. E., Sarajedini, A., Geisler, D., Gallart, C., \& Wischnjewsky, M. 2007a, MNRAS, 381, L84
\bibitem[Piatti et al.(2007b)]{Piatti_07b}
Piatti, A. E., Sarajedini, A., Geisler, D., Gallart, C., \& Wischnjewsky, M. 2007b, MNRAS, 382, 1203
\bibitem[Pozzetti et al.(2007)]{Pozzetti_07}
Pozzetti, L. et al. 2007, A\&A, 474, 443
\bibitem[Rolleston et al.(2003)]{Rolleston_03}
Rolleston, W. R. J., Venn, K., Tolstoy, E., \& Dufton, P. L. 2003, A\&A, 400, 21
\bibitem[Springel et al.(2005)]{Springel_05}
Springel, V. et al. 2005, Nature, 435, 629
\bibitem[Stanimirovi\'{c} et al.(2004)]{Stanimirovic_04}
Stanimirovi\'{c}, S., Staveley-Smith, L., \& Jones, P. A. 2004, ApJ, 604, 176
\bibitem[Tsujimoto et al.(1995)]{Tsujimoto_95}
Tsujimoto, T., Nomoto, K., Yoshii, Y., Hashimoto, M., Yanagida, S., \& Thielemann, F.-K. 1995, MNRAS, 277, 945
\bibitem[Tsujimoto(2007)]{Tsujimoto_07}
Tsujimoto, T. 2007, ApJ, 665, L115
\bibitem[van der Marel et al. (2002)]{vandermarel_02}
van der Marel, R. P., Alves, D. R., Hardy, E., \& Suntzeff, N. B. 2002, AJ, 124, 2639
\bibitem[Venn(1999)]{Venn_99}
Venn, K. A. 1999, ApJ, 518, 405
\bibitem[Yoshii et al.(1996)]{Yoshii_96}
Yoshii, Y., Tsujimoto, T., \& Nomoto, K. 1996, ApJ, 462, 266
\bibitem[White \& Rees(1978)]{White_78}
White, S. D. M., \& Rees, M. J. 1978, MNRAS, 183, 341
\bibitem[White \& Frenk(1991)]{White_91}
White, S. D. M., \& Frenk, C. S. 1991, ApJ, 379, 52
\end{thebibliography}
\end{document}